\documentstyle[preprint,pra,aps]{revtex}
\tightenlines
\begin{document}
\draft
\title{\bf Off-Diagonal Hyperfine Interaction and Parity Non-conservation 
in Cesium}
\author{V.A.Dzuba and V.V.Flambaum}
\address{School of Physics, University of New South Wales, 
Sydney 2052,Australia}
\date{\today}
\maketitle

\begin{abstract}
We have performed relativistic many-body calculations of the hyperfine 
interaction in the $6s$ and $7s$ states of Cs, including the off-diagonal
matrix element. The calculations were used to determine the accuracy of the 
semi-empirical formula for the electromagnetic transition amplitude
$\langle 6s|M1|7s \rangle$
 induced by the hyperfine interaction. We have found that even though the
contribution of the many-body effects into  the 
matrix elements is very large, the square root formula 
$\langle 6s|H_{hfs}|7s \rangle = \sqrt{\langle 6s|H_{hfs}|6s \rangle
\langle 7s|H_{hfs}|7s \rangle}$
remains valid to the accuracy of a fraction of $10^{-3}$. 
The result for the M1-amplitude is used
in the interpretation of the  parity-violation measurement
in the $6s-7s$ transition in Cs which claims a possible deviation from the 
Standard model.

\end{abstract} 
\vspace{1cm}


\section{Introduction}

Recent progress in highly accurate measurements of parity non-conservation
(PNC) in atoms has got to the point where new physics beyond the Standard 
Model of elementary particles can  be studied.
The latest analysis \cite{Bennett}
of the most precise measurements of the PNC in cesium \cite{Wood}
suggests that the value of the weak charge of the $^{133}$Cs nucleus may 
differ from the prediction of the Standard Model.
In that experiment \cite{Wood} the ratio of the PNC $E1$ amplitude to
the tensor polarizability $\beta$ for the $7S_{1/2} - 6S_{1/2}$ transition
was measured with 0.35\% accuracy. The measured value can be written in the 
form
\begin{equation}
	\frac{k_{PNC}}{\beta}\frac{Q_W}{N},
\label{PNC}
\end{equation}
where $k_{PNC}$ is the electron matrix element of the electric dipole
transition induced by the weak interaction between $7S_{1/2}$ and $6S_{1/2}$
states of $^{133}$Cs, $Q_W$ is the weak nuclear charge and N is the number
of neutrons. To interpret the measurements in terms of the weak nuclear
charge one needs to know $k_{PNC}$ and $\beta$. The value of $k_{PNC}$ can 
be obtained from atomic calculations only. Bennett and Wieman \cite{Bennett} 
used the value $k_{\rm PNC} = 0.9065(36) iea_0$ which is the average of our 
result $k_{\rm PNC} = 0.908(9) iea_0$ \cite{Dzuba89a} obtained in 1989 and 
the result of the Notre-Dame group $k_{\rm PNC} = 0.905(9) iea_0$ 
\cite{Blundell} obtained in 1990. Note that Bennett and Wieman
assumed 0.4\% accuracy of the calculations contrary to the 1\% accuracy
claimed in both calculations. This assumption was based on the 
comparison of the calculated atomic quantities relevant to the PNC
amplitude (electromagnetic transition amplitudes between lower $s$ and
 $p$ states
and hyperfine structure intervals of these states) with the latest very 
accurate measurements which resolved 
major discrepancies between theory and experiment in favor of theory.

The most precise value of $\beta, \beta = 27.024(43)(67)a_0^3$,
was obtained in Ref. \cite{Bennett} 
from the measurements of the ratio $M1_{hfs}/\beta$ where $M1_{hfs}$ 
is the $M1$ transition amplitude between the states $6S$ and $7S$ induced by
the hyperfine structure (hfs) interaction.
Semiempirical formula for the $M1_{hfs}$ amplitude derived in Refs.
\cite{Guena84,Piketty,Guena88} was used in the analysis:
\begin{equation}
	M1_{hfs} = - \left|\frac{\mu_B}{c}\right|
	\frac{\sqrt{A_{6s}A_{7s}}}{E_{7s}-E_{6s}}
	\frac{1}{2}(g_S - g_I) 1.0024
\label{m1hfs}
\end{equation}
Here $A_{6s}$ and $A_{7s}$ are the hfs constants of the $6s$ and $7s$ states 
of Cs, $g_S = 2.0025, g_I = -0.0004$,
the coefficient 1.0024 was introduced to account for the many-body 
effects.
This gives $M1_{hfs} =  |\frac{\mu_B}{c}| 0.8094(20) \times 10^{-5}$
\cite{Piketty,Guena88}.

Values $\beta = 27.024(43)(67)a_0^3$ and $k_{PNC} = 0.9065(36) iea_0$
and measurements of (\ref{PNC}) \cite{Wood} lead to the value of the 
weak charge of $^{133}$Cs $Q_W = -72.06(28)(34)$ which differs 
from the prediction of the Standard Model $Q_W = -73.20(13)$ \cite{Marciano}
by $2.5\sigma$.

From the point of view of accurate atomic calculations, there are two 
major questions in the analysis above which should be considered.
The first is whether the actual accuracy of the PNC calculations is really
0.4\%. The second is whether the semi-empirical formula (\ref{m1hfs}) is
accurate.
In the present paper we address the second question, leaving the first 
one for later work.

\section{Preliminary Analysis}

$M1_{hfs}$ amplitude appears due to mixing of the $6s$ and $7s$ states
by the hfs interaction,
\begin{eqnarray}
	M1_{hfs} &=& \frac{\langle 6s,F |H_{hfs}| 7s,F \rangle}
	{E_{6s} - E_{7s}}\langle 7s,F |M1| 7s,F' \rangle  \nonumber \\
	&+& \langle 6s,F |M1| 6s,F' \rangle
	\frac{\langle 6s,F' |H_{hfs}| 7s,F' \rangle}{E_{7s} - E_{6s}}.
\label{m1}
\end{eqnarray}
Two major assumptions have been made to arrive at (\ref{m1hfs}) 
from (\ref{m1}).
First, the non-relativistic expression for the operator of the $M1$ 
transition was used:
\begin{equation}
	M1 = -|\mu_B|g ({\mathbf L +2 S}).
\label{m1o}
\end{equation}
Second, the square root formula is assumed to be valid
\begin{equation}
	\langle 6s |H_{hfs}| 7s \rangle =
	\sqrt{\langle 6s|H_{hfs}|6s \rangle \langle 7s|H_{hfs}|7s \rangle}.
\label{AA}
\end{equation}
The accuracy of both of these assumptions needs to be examined. The situation
is  clear with the relativistic corrections to the $M1$ operator
(\ref{m1o}). According to the estimations of Bouchiat
and Piketty \cite{Piketty} the relativistic effects modify the 
amplitudes $\langle 6s |M1| 6s \rangle$ and
$\langle 7s |M1| 7s \rangle$ at only the $10^{-4}$ level. This is in line
 with the many-body calculations
of the relativistic effects in $g$-factors and $M1$-transition amplitudes for
Cs and other  alkaline atoms in our early works \cite{Flambaum,Dzuba85}.

The situation with the square root formula (\ref{AA}) is less clear.
In their pioneering work Bouchiat and Piketty \cite{Piketty} estimated 
the first order core 
polarization corrections to it and introduced the correction factor 1.0017.
In a later paper by  Bouchiat and Gu\'{e}na \cite{Guena88} this factor
was assumed to be 1.0024 (see also formula (\ref{m1hfs})).
The accuracy of the estimation of the many body correction was assumed
to be approximately equal to the correction itself ($\sim 0.002$) 
\cite{Piketty,Guena88}.
In these works there were no accurate calculations of other many body 
contributions to the hfs beyond the first order core polarization
corrections. However, it is known that these contributions can be
up to 20\% of the hyperfine structure (see below).
The applicability of eq. (\ref{m1hfs}) in this situation is not
obvious.

The accurate relativistic many-body calculations of the off-diagonal
 hfs matrix element
(\ref{AA}) were recently performed by the Notre Dame group \cite{Derevianko}.
The accuracy of the calculations was about 1\% and agreement with formula 
(\ref{AA}) within this accuracy was achieved. Note that the theoretical 
accuracy for the diagonal hfs matrix elements is also about 1\% (see Refs. 
\cite{Dzuba89b,Safronova} and this article).
This accuracy is not sufficient to find an accurate value
of $\beta$ to add anything new to the result of
the cesium PNC experiment published in \cite{Wood}.

However, we believe that the validity of the square root formula (\ref{AA})
can be demonstrated to much higher accuracy than the absolute theoretical
accuracy of the hfs calculations
(here we agree with \cite{Guena84,Piketty,Guena88}).
We suggest that the following combination of matrix elements be calculated
\begin{equation}
	R = \frac{\langle 6s |H_{hfs}| 7s \rangle}
	{\sqrt{\langle 6s |H_{hfs}| 6s \rangle 
	\langle 7s |H_{hfs}| 7s \rangle}} - 1,
\label{main}
\end{equation}
where all hfs  matrix elements are calculated in the same 
approximation. The value of $R$ can be calculated with very high accuracy
because uncertainties in different matrix elements cancel each other
almost exactly. We will demonstrate that inclusion of different many body
and relativistic effects leave the formula
\begin{equation}
	\langle 6s |H_{hfs}| 7s \rangle =
	\sqrt{\langle 6s |H_{hfs}| 6s \rangle 
	\langle 7s |H_{hfs}| 7s \rangle}
\label{main1}
\end{equation}
valid to very high accuracy, so that the value of $R$ (\ref{main}) remains
very small.

Let us start from the analytical estimates of different contributions 
to $R$ in (\ref{main}).
First note that in the single-electron approximation formula (\ref{main1})
is exact if the wave functions of the $6s$ and $7s$ states are 
proportional 
\begin{equation}
	\psi_{6s} = B \psi_{7s}
\label{pro}
\end{equation}
($a_0$ is Bohr radius)
on short distances from the nucleus, $r \leq a_0/Z$. Dirac equations for the
states $6s$ and $7s$ differ by the energy only. Therefore,
 their solutions on short
distances where the difference in energies is small compared to the potential,
differ by normalization only. 
One can say that (\ref{pro}) is valid if
\begin{equation}
	\Delta E/|V| \ll 1,
\label{ll} 
\end{equation}
where $\Delta E = 0.08445$ a.u. is the energy difference 
between the $6s$ and $7s$ states of Cs, $V$ is the atomic potential.
The Hamiltonian of the hfs interaction $H_{hfs}$ is proportional to $1/r^3$ 
and the main contribution to its matrix elements comes from  the distances 
$r \leq a_0/Z$.
Substitution of $V = Ze^2/r, r = a_0/Z$ and $Z=55$ into (\ref{ll}) gives
\begin{equation}
	\frac{\Delta E}{V} \approx 3 \times 10^{-5}.
\label{DEV} 
\end{equation}
Note that for $s$-waves the correction can be even smaller. Indeed,
in the non-relativistic approximation $s$-wave hfs is proportional
to $\delta(r)$. Thus, the typical distances 
$r \sim \hbar/(m_e c) = \alpha a_0$, where $ \alpha = 1/137$.

Let us now consider the many-body effects. It is convenient to do this
using the many body perturbation theory in the residual Coulomb interaction 
$U, \ U = H - H_{HF}$.
Here $H$ is the exact Hamiltonian of the atom and $H_{HF}$ is the
Hartree-Fock Hamiltonian. We generate the complete zero-approximation
set of the eigenvalues, wave functions and Green's functions
using the Hartree-Fock Hamiltonian. The small parameter of this many-body
perturbation theory is the ratio of the non-diagonal matrix element
of the residual interaction $U$ to the large energy denominator for excitation
of the electron from the closed electron shell (electron core),
 e.g. 5p -electron: $U/E_{5p} \sim10^{-2}$.
 
  The perturbative (correlation) corrections to the hfs matrix element
can be divided into two classes:
 the self-energy corrections and the vertex 
corrections. The former can be included into eq. (\ref{main}) through the 
redefinition of the single electron wave functions while the latter
are included through the redefinition of the $H_{hfs}$ operator. 

Self-energy corrections dominate in the hfs of alkaline atoms
(see, e.g. \cite{Dzuba84}). The major contribution is due to
the correlations between an external electron and core electrons.
We include them by using so called  Brueckner orbitals  instead of the 
Hartree-Fock orbitals as the single-electron wave functions in eq.
 (\ref{main}).
The Brueckner orbitals are obtained by introducing an additional operator
$\hat \Sigma$ into the  Hartree-Fock equations for the external electron
and solving the Dyson-type equation $(H_{HF} + \hat \Sigma(E) - E)\psi = 0$. 
The $\hat \Sigma$ is an energy-dependent non-local
operator which is also called the ``correlation potential'' 
\cite{Dzuba89b,Dzuba87}. For the calculation of $\hat \Sigma$ see the next
section.
The Brueckner type correlation correction constitutes 20\% of the hfs of 
$6s$ and $7s$ states of Cs. However, if we neglect the dependence of 
$\hat \Sigma$ on energy, the estimation (\ref{DEV})
is still valid. It follows from the calculations that 
$\partial \hat \Sigma/\partial E \sim 1\%$ for $E \sim E_{6s},E_{7s}$
 (it is suppressed by the parameter $ \Delta E/E_{5p}$). This
leaves condition (\ref{DEV}) practically unchanged.

Dominating vertex corrections to the hfs matrix element are due to the
effect of core polarization by the nuclear dipole magnetic field.
Since the core states change in the magnetic field, 
the Hartree-Fock potential $V$ created by the core electrons as well as 
the correlation potential $\hat \Sigma$ also change. The effect of this 
change on the hfs can be accounted for by redefining the operator of the 
hfs interaction 
\cite{Dzuba89b}:
\begin{equation}
	H'_{hfs} = H_{hfs} + \delta V + \delta \hat \Sigma.
\label{Hhfs} 
\end{equation}
The correction to the hfs caused by $\delta V$ is often called the 
RPA-type \cite{RPA}
correction, while another correction associated with $\delta \hat \Sigma$ 
is the non-Brueckner correlation correction or structural radiation 
\cite{Dzuba87}.
These corrections are more likely to cause deviation from the square root
formula since they are localized on larger distances up to the core radius.
Note, however, that in the case of the hfs interaction $\delta V$ is completely
due to the Hartree-Fock exchange potential. There is no change to the 
Hartree-Fock direct potential since magnetic field does not
change electron density in the first order of perturbation theory.
This means that $\delta V$ vanishes exponentially outside the core.
 Inside the core, at $r \sim  a_0$ , $\Delta E/V \sim 0.01$ and the
$6s$ and $7s$ orbitals are still proportional. Note that the potential $V$
at these distances may be estimated as $V \sim - Z_{eff} e^2/r$,  where
 $Z_{eff} \sim 5$. Since the contribution of  $\delta V$ is about 
10\% we come to the estimate $10^{-3}$ for the error of the square
 root formula.

There is one more reason why the square root formula is  accurate.
The expression for $R$ (\ref{main})
is symmetric with respect to the energies $E_{6s}$ and $E_{7s}$. Therefore,
its decomposition over $\Delta E$ ($\Delta E = E_{6s} - E_{7s}$) starts
from $\Delta E^2$:
\begin{equation}
	R = a(\Delta E)^2 + b(\Delta E)^4 + \ldots.
\label{R} 
\end{equation}
Since all linear in $\Delta E$ terms are canceled out one can say that
the error should be smaller than in the estimates above.
The dimensionless parameter for (\ref{R}) is
\[
	(\Delta E/E_{5p})^2 \sim 10^{-2},
\]
where $E_{5p} \approx 0.84$ a.u. is the core excitation energy. 
Since the term $a (\Delta E)^2$ arises due to $\delta V$ and 
$\delta \hat \Sigma$ which contribute about 10\% and 1\%, respectively, 
into the hfs, the total deviation from the
square root formula caused by the RPA and non-Brueckner corrections
should be smaller than  $10^{-1} \times 10^{-2} = 10^{-3}$.
We may add that the contribution of $\delta \hat \Sigma$ to the hfs is 
much smaller than the contribution of $\hat \Sigma$ since 
$\delta \hat \Sigma$ has an additional
suppression by the parameter $(\Delta E/E_{5p})$ \cite{Dzuba87}.

There are also contributions to the self-energy and vertex due to the radiative
corrections. We have not considered these contributions in our calculations.
 However, they come from the very short distances
$r \leq \hbar/m_ec = \alpha a_0$ and should not cause any significant
deviation from the square root formula. 

Finally, let us estimate contributions to $M1_{hfs}$ which cannot be
presented in the form of eq. (\ref{m1}). Let us use the basis of the exact
atomic eigenstates and treat $H_{hfs}$ as a perturbation in this basis.
The result can be presented in the form
\begin{eqnarray}
	M1_{hfs}& = & \sum_{\alpha}\frac{\langle \widetilde{6s},F|H_{hfs}
	|\alpha,F \rangle}{E_{6s} - E_{\alpha}}
	\langle \alpha,F|M1|\widetilde{7s},F' \rangle \nonumber \\
	 &+&   \sum_{\beta}\langle \widetilde{6s},F|M1|\beta,F' \rangle 
	\frac{\langle \beta,F' |H_{hfs}|\widetilde{7s},F' \rangle}
	{E_{\beta} - E_{7s}}.
\label{nd}
\end{eqnarray}
Here $|\widetilde{6s} \rangle, |\widetilde{7s} \rangle, |\alpha \rangle,
|\beta \rangle$ are the eigenstates which include all possible configuration
mixing, $|\alpha \rangle$ and $|\beta \rangle$ may contain 
an arbitrary number of pairs of excited electrons and holes in the electron 
core. All non-diagonal
matrix elements of the $M1$ operator vanish in the non-relativistic limit. 
Moreover,  it was demonstrated in our work \cite{Flambaum} that 
the dominant contribution appears only in the second order in the
spin-orbit interaction and in the first order in configuration mixing, i.e. 
non-diagonal $M1$ matrix elements are of the order 
$M1 \sim (Z\alpha)^4 Q_{in}/E \sim 10^{-4}-10^{-5} |\mu_B|$, where $Q_{in}$ 
is the non-diagonal matrix element of the Coulomb interaction corresponding
to an excitation of a core electron and $E$ is the energy of this excitation.
Indeed, the operator of the magnetic moment is
$M1 = \mu_B ({\mathbf L+2S}) =\mu_B ({\mathbf 2J-L})$
(relativistic correction to this expression $\sim 10^{-5}$).
Electron wave functions are the eigenfunctions of the total electron
angular momentum ${\mathbf J}$. Therefore, ${\mathbf J}$ does not give
any non-diagonal matrix elements. On the other hand, the matrix element
$\langle \alpha,J=1/2|{\mathbf L}|\beta,J=1/2 \rangle$ requires
spin-orbit interaction both in the bra $\langle \alpha,J=1/2|$ and
ket $|\beta,J=1/2 \rangle$ vectors, since in the non-relativistic limit
they correspond to the total orbital angular momentum $L=0$,
i.e. ${\mathbf L}|\alpha \rangle = {\mathbf L}|\beta \rangle = 0$.
Thus, we need the second order in spin-orbit interaction.
Note that the non-diagonal in angular momentum $L$ matrix elements of the 
hyperfine interaction like 
$\langle \tilde s_{1/2}|H_{hfs}|\tilde d_{3/2} \rangle$ do not help
since in this case both the hfs matrix element and $M1$ matrix element
$\langle \tilde s_{1/2}|M1|\tilde d_{3/2} \rangle$ are very small.

Non-diagonal matrix elements of $M1$ were calculated in Refs.
\cite{Flambaum,Dzuba85}; the value
$\langle 6s|M1|7s \rangle \approx 0.4 \times 10^{-4} |\mu_B|$ was
measured in Refs. \cite{Hoffnagle,Guena84,Gilbert}.
Thus, each term with the non-diagonal $M1$ matrix element in eq. (\ref{nd})
is suppressed by a factor of $10^{-4} - 10^{-5}$. Therefore, we may
safely assume that the correction to the diagonal $M1$ contribution
(\ref{m1}) does not exceed $10^{-3}$.

We should note that it may not be easy to come to this conclusion using
perturbation theory in the Dirac basis of electron orbitals ($jj$ scheme)
(Dirac basis was used in Ref. \cite{Piketty}). In this basis the small result
must appear due to strong cancellations between different terms in the sum
over intermediate states.

\section{Many-Body Calculations}

To test the validity of the square root formula (\ref{main1}) we performed  
accurate many-body 
relativistic calculations of the off-diagonal and diagonal hfs matrix elements.
Detailed discussion of the accurate hfs calculations can be found elsewhere
\cite{Dzuba89b}. Here we repeat the main points emphasizing the role of
 different many-body effects.

\begin{figure}[b]
\psfig{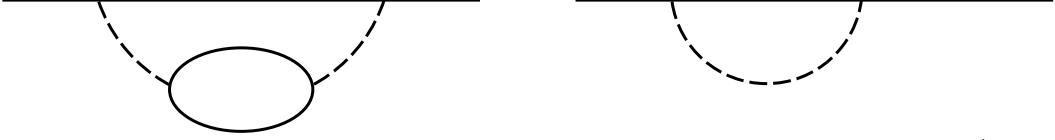}
\caption{Second-order diagrams for the self-energy of the valence 
electron ($\hat \Sigma$ operator). Dashed line is the Coulomb interaction 
between core and valence electrons. Loop is the polarization of the atomic
core which corresponds to the virtual creation of the excited electron and
a hole in the core shells.}
\label{sigma}
\end{figure}

We start calculations from the relativistic Hartree-Fock (RHF) method
in the $V^{N-1}$ approximation (calculations for the external electron
are carried out in the frozen self-consistent field of the core  ).
The core polarization is calculated using the Hartree-Fock equations
in an external field  \cite{Dzuba84}. It is equivalent to the well-known 
random-phase approximation with exchange method (see, e.g. \cite{RPA}).
The  many-body effects such as the  Brueckner-type correlations,
and the structural radiation are included by means of the
correlation potential method \cite{Dzuba87}.
As it was pointed out in the previous section, the Brueckner-type correlation
corrections are included by solving the Dyson-type equation for the states
of the external electron
\begin{equation}
	(H_{HF} + \hat \Sigma - E) \psi = 0.
\label{Bru}
\end{equation}
Correlation potential $\hat \Sigma$ 
accounts for the correlation between an external electron and core
electrons. We use many body perturbation theory and the Feynman diagram
technique to calculate $\hat \Sigma$ \cite{Dzuba89b,Dzuba89c}. 
The perturbation expansion of $\hat \Sigma$
in the residual Coulomb interaction starts from the second order. 
The corresponding diagrams are presented on Fig. \ref{sigma}.
We include both the second order diagrams and three dominating classes of 
higher order correlations:
\begin{enumerate}
\item Screening of the Coulomb interaction between an external electron 
and core electrons by other core electrons. This is a collective
phenomenon and the corresponding chain of diagrams is enhanced by a
factor approximately equal to the number of electrons in the external
closed subshell (the $5p$ electrons on Cs). We stress that our approach 
takes into account screening diagrams with double, triple and higher core
electron excitations in contrast to the popular coupled cluster method
where only double and selected triple excitations are considered
(see, e.g. \cite{Safronova}). The effect of screening is taken into 
account in all orders by  summation of the corresponding
chain of diagrams which in the Feynman digram technique form a matrix 
geometrical progression.
\item Hole-particle interaction in the core polarization operator.
This effect is enhanced by the large zero-multipolarity diagonal
matrix elements of the Coulomb interaction.
We take it into account by amending the direct Hartree-Fock potential
in which the polarization operator is calculated.
\item Iterations of the self-energy operator ($\hat \Sigma$).
This chain of diagrams describes the nonlinear effects of the 
correlation potential and is enhanced by the small denominator,
which is the excitation energy of an external electron (in comparison
with the excitation energy of a core electron). The iterations of $\hat \Sigma$
are included by solving equation (\ref{Bru}).
\end{enumerate}

\begin{figure}[b]
\psfig{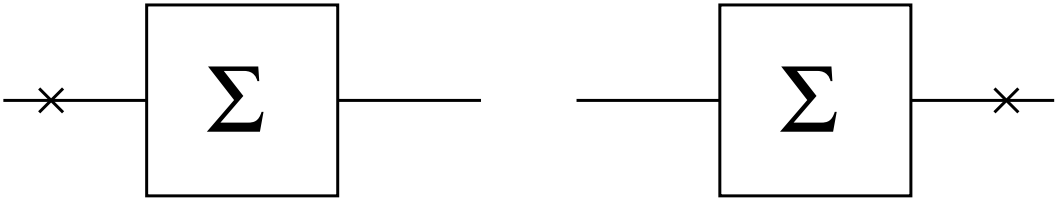}
\caption{Bruckner-type correlation diagrams for the hfs. Cross denotes
the hfs interaction. The $\Sigma$ operator includes second-order diagrams 
(Fig.1) and higher-order diagrams as described in the text.}
\label{bru}
\end{figure}

Substituting the Brueckner orbitals into
eq. (\ref{main}) accounts for the dominating correlation corrections
to the hfs. Corresponding diagrams are presented on Fig. \ref{bru}.
These corrections constitutes 23\%
of the hfs of the $6s$ state of Cs and 12\% of the hfs of the $7s$ state
of Cs.

\begin{figure}[b]
\psfig{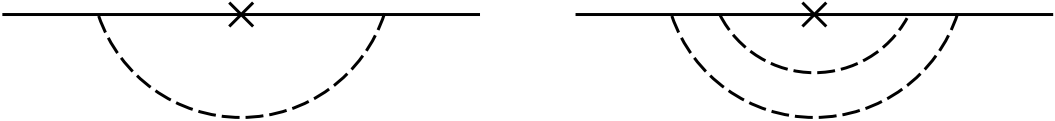}
\caption{Core polarization (RPA) diagrams for the hfs in the first and 
second order in Coulomb interaction.}
\label{rpae}
\end{figure}

To take into account the core polarization effect we self-consistently solve
the Hartree-Fock equation for the core states in the nuclear magnetic field.
The details are presented in Ref. \cite{Dzuba84}.
When all corrections $\delta \psi_n$ to core states caused by the 
magnetic field are found, they are
used to calculate the correction $\delta V$ to the Hartree-Fock 
potential. Then core polarization is included into the single-electron
matrix element $\langle a|H_{hfs}|b \rangle$ between valence states
$|a \rangle$ and $|b \rangle$ by redefining the operator of the hfs
interaction $ H'_{hfs} = H_{hfs} + \delta V$.
This corresponds to the summation of the infinite series of the RPA-type
of diagrams presented on Fig. \ref{rpae}.
The RPA-type core polarization contribution to the hfs of the $6s$ and $7s$
states of Cs is about 15\%.

\begin{figure}[b]
\psfig{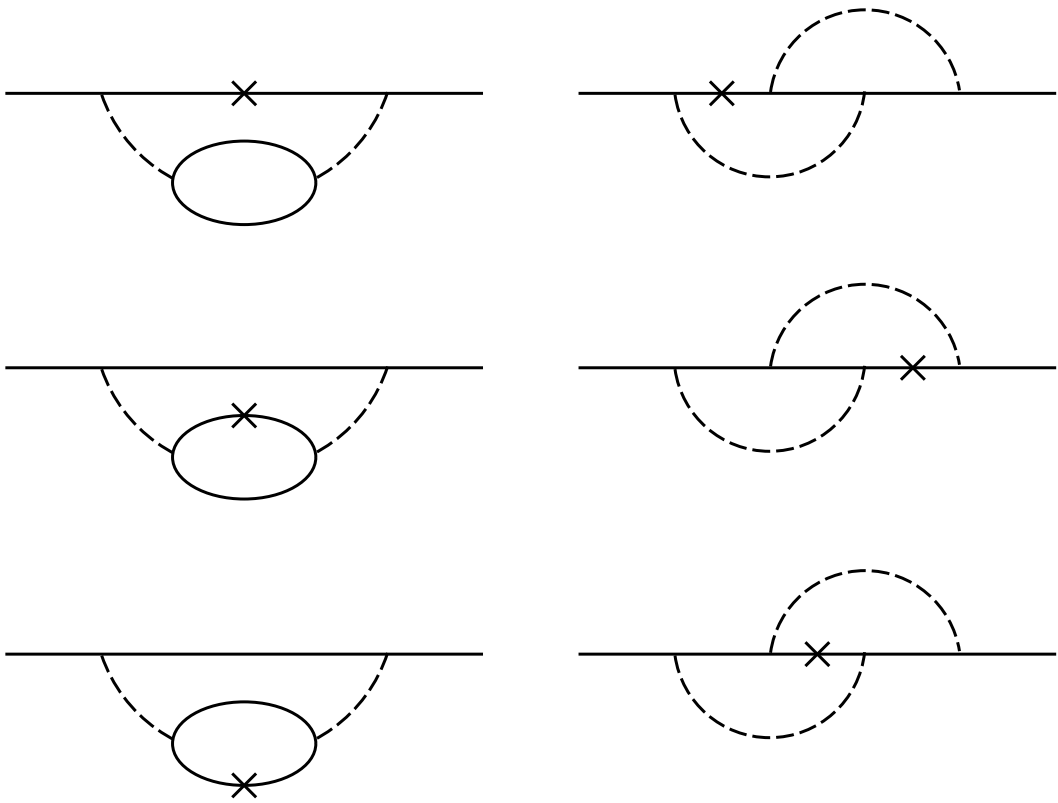}
\caption{Structural radiation}
\label{str}
\end{figure}

Core polarization also leads to the change of $\hat \Sigma$. Corresponding
contributions to the hfs matrix element 
$\langle a|\delta \hat \Sigma|b \rangle$
are often called structural radiation. 
Second order diagrams for structure radiation are presented 
on Fig. \ref{str}. 
We use direct summation over the complete set of single-electron 
states to calculate these diagrams.

There is also a contribution to the hfs due to the change of normalization
of the wave function caused by $\hat \Sigma$. This contribution can be 
written in a form 
\begin{equation}
	A_{norm} = \frac{1}{2}\langle a|H_{hfs}|b \rangle 
	(\langle a |\partial \hat \Sigma/\partial E|a \rangle +
	 \langle b |\partial \hat \Sigma/\partial E|b \rangle).
\label{norm}
\end{equation}
The combined contribution of the structural radiation and renormalization
into the hfs of the $6s$ and $7s$ states of Cs are 1.5\% and 0.6\% 
respectively.

When all dominating higher-order correlations are included into the 
calculation of the Brueckner orbitals for the  $6s$ and $7s$ states of cesium, 
the accuracy for the calculated energies of these states is very high and 
constitutes about 0.1\%. However, we introduced fitting parameters to re-scale
$\hat \Sigma$ to fit the energies exactly. This procedure allows us to 
effectively include some omitted higher order correlations and to test 
the sensitivity of the hfs matrix elements on the value of $\hat \Sigma$.

The results for the hfs are presented in Table \ref{tab}.
In this table $h \equiv H_{hfs}$, the matrix elements of $\delta V$
are RPA-type corrections, the matrix elements
of $\delta \hat \Sigma$ are structural radiation (including renormalization 
(\ref{norm})), matrix elements with $\psi_{Br}$ include Brueckner-type
correlation corrections.
One can see that the correction  to the square root formula
due to the considered many-body effects does not exceed $4.4 \times 10^{-4}$.
When all dominating many-body effects are taken into account the accuracy
of the calculated hfs constants compared to experiment is about 1\%.
However the square root formula is still valid to the accuracy of about 
$10^{-4}$. The most likely cause of the remaining discrepancy with 
experiment is higher-order correlation corrections not included in
our calculations. These corrections are localized
on the radius of the core and due to the fact that these corrections are
 very small ($\sim$ 1\% of the
experimental hfs) it is extremely unlikely that they can break the
square root formula. The same may be said about the very small radiative
and Breit corrections.
Note that our final results for the diagonal hfs matrix elements for
the $6s$ and $7s$ states are in very good agreement with the calculations
of the Notre-Dame group \cite{Safronova}. 

It follows from the above that the correction to the square root formula
is about an order of magnitude smaller than the estimations of Bouchiat
and Gu\'{e}na \cite{Guena88}. However, there is no formal disagreement between
the results, since  Bouchiat and Gu\'{e}na estimated the uncertainty
of their result to be equal to the correction itself.
We believe that for the analysis of the PNC experiment it is safer to
assume no correction to the square root formula. This slightly changes the
numbers. The $M1_{hfs}$ amplitude, tensor polarizability $\beta$ and
weak charge of the $^{133}$Cs nucleus become
\begin{eqnarray}
	M1_{hfs}& = & |\frac{\mu_B}{c}| 0.8074(8) \times 10^{-5},\nonumber \\
	\beta  & = & 26.957(43)(27) a_0^3, \label{Mres}\\
	Q_W    & = & -71.88(28)(29). \nonumber
\end{eqnarray}
To stress the importance of the result here we used an estimate of
the theoretical accuracy  0.4\% \cite{Bennett} in the value of  $k_{PNC}$.
Our result for $M1_{hfs}$ is in very good agreement with the result of
Derevianko {\it et al} \cite{Derevianko}
\begin{eqnarray}
	M1_{hfs}& = & |\frac{\mu_B}{c}| 0.8070(73) \times 10^{-5},
\label{Derevo}
\end{eqnarray}
but has the better accuracy.
The weak nuclear charge $Q_W$ in (\ref{Mres}) represents even larger 
deviation from the Standard Model value $Q_W = -73.20(13)$ \cite{Marciano}
 than the 
result presented by Bennett and Wieman \cite{Bennett}. 
The deviation is $2.9\sigma$ if 0.4\% accuracy of calculations of
the $k_{PNC}$ is assumed. Note that even if 1\% accuracy is assumed
for the calculated value of $k_{PNC}$ as it was claimed in both 
theoretical works \cite{Dzuba89a,Blundell} then there is still 
$1.5\sigma$ deviation from the Standard Model.
However, we would like to stress once more that before making any conclusions
about agreement or disagreement with the Standard Model the question 
about the accuracy of the atomic calculations of the PNC electronic
matrix element $k_{PNC}$ (see (\ref{PNC})) should be carefully re-analyzed.


\begin{table}
\caption{Hyperfine structure matrix elements for the 6S and 7S states
of $^{133}$Cs (MHz).}
\label{tab}
\begin{tabular}{llccccc}
 \multicolumn{2}{c}{Approximation} & $A_{6S}$ & $A_{7S}$ &
 $\sqrt{A_{6S} A_{7S}}$ & $ \langle 6S |h| 7S \rangle $ &
 $ \frac{\langle 6S |h| 7S \rangle }{\sqrt{A_{6S} A_{7S}}} - 1$ \\
\hline
 $ \psi = \psi_{HF}$ & $ \langle \psi |h| \psi \rangle $
 &  \dec 1424.8 & \dec  391.5 & \dec  746.9 & \dec  746.9 & 0\\
 & $ \langle \psi |h + \delta V| \psi \rangle $
 & \dec 1712.5 & \dec  469.7 & \dec  896.9 & \dec 897.1 & $2.2\times 10^{-4}$\\
 & $ \langle \psi |h + \delta V + \delta \hat \Sigma| \psi \rangle $
 & \dec 1687.8 & \dec  466.9 & \dec  887.7 & \dec 887.6 & $1.1\times 10^{-4}$\\
\hline
 $ \psi = \psi_{Br}$ &$ \langle \psi |h| \psi \rangle $
 & \dec 1952.4 & \dec  459.5 & \dec  947.2 & \dec  947.2 & 0\\
 & $ \langle \psi |h + \delta V| \psi \rangle $
 & \dec 2302.0 & \dec  541.4 & \dec 1116.3 &\dec 1116.7 & $3.5\times 10^{-4}$\\
 & $ \langle \psi |h + \delta V + \delta \hat \Sigma| \psi \rangle $
 & \dec 2267.6 & \dec  537.7 & \dec 1104.3 &\dec 1104.5 & $1.8\times 10^{-4}$\\
\hline
 $ \psi = \psi_{fit}$\tablenotemark[1]
 & $ \langle \psi |h + \delta V| \psi \rangle $
 & \dec 2308.3 & \dec  542.5 & \dec 1119.0 &\dec 1119.5 & $4.4\times 10^{-4}$\\
 & $ \langle \psi |h + \delta V + \delta \hat \Sigma| \psi \rangle $
 & \dec 2273.8 & \dec  538.8 & \dec 1106.9 &\dec 1107.3 & $3.6\times 10^{-4}$\\
\hline
 \multicolumn{2}{c}{SDpT\tablenotemark[2]}
 & \dec 2278.5 & \dec  540.6 & \dec 1109.8 &    &  \\
\hline
 \multicolumn{2}{c}{Experiment\tablenotemark[3]}
 & \dec 2298.2 & \dec  545.9 & \dec 1120.1 &    &  \\
\end{tabular}
\tablenotetext[1]{Brueckner orbitals with $\hat \Sigma$ operator 
rescaled to fit the energy.}
\tablenotetext[2]{Single, double and partly triple excitation approximation;
calculations by the Notre-Dame group, reference \cite{Safronova}}
\tablenotetext[3]{Reference \cite{hfs}}
\end{table}

\widetext
\newpage
\input psfig
\psfull

\end{document}